\documentclass[twocolumn,         
               showpacs,longbibliography,            
               showkeys,preprintnumbers,     
               aps,                 
               prd,                 
               letterpaper,             
               superscriptaddress,      
               nofootinbib,         
               tightenlines,        
               floats,floatfix      
               ]{revtex4-1}
\usepackage{physics}
\usepackage{epsf}
\usepackage{graphicx,color}
\usepackage{subfigure}
\usepackage{latexsym}
\usepackage{amsmath,amssymb}
\usepackage[colorlinks=true,linkcolor=blue,citecolor=blue]{hyperref}
\usepackage{mathrsfs}
\usepackage{comment}
\usepackage{soul}
\usepackage{cancel}
\definecolor{purple}{rgb}{0.58,0.0,0.83}
\definecolor{orange}{rgb}{1,0.5,0}
\DeclareSymbolFontAlphabet{\mathrsfs}{rsfs}
\DeclareMathAlphabet{\mathcal}{OMS}{cmsy}{m}{n}

\begin{document}

\title{Temporal Correlations Between Fuzzy Dark Matter and Baryonic Matter in Virialized Core--Halo Structures}

\author{Iv\'an \'Alvarez-Rios}
\email{ivan.alvarez@umich.mx}

\author{Francisco S. Guzmán}
\email{francisco.s.guzman@umich.mx}

\affiliation{
Instituto de F\'isica y Matem\'aticas, Universidad Michoacana de San Nicol\'as de Hidalgo, 
Edificio C-3, Ciudad Universitaria, 58040 Morelia, Michoac\'an, M\'exico
}

\date{\today}

\begin{abstract}
Fuzzy Dark Matter (FDM) predicts the existence of virialized halos with interference-driven granular structure generated by the wave nature of ultralight bosons. Since these fluctuations produce time-dependent gravitational fields, they can be proposed as a potential source of observable dynamical effects on baryonic matter. In this work we test whether such fluctuations generate measurable temporal correlations between FDM and a coupled gas component. We evolve coupled FDM-gas configurations with the Schr\"odinger-Poisson-Euler system in three dimensions. The FDM component is initialized through multimerger configurations that relax toward core-halo structures, while the baryonic component is modeled as an ideal gas. After virialization, we analyze the temporal fluctuations of both components through correlation functions and spectral diagnostics. The FDM density field shows fluctuations with significant high-frequency content, whereas the gas develops smoother and longer-lived coherent structures. Although the gas dynamically responds to the gravitational potential generated by the FDM halo, we find no strong temporal correlation between the interference-driven fluctuations of FDM and the gas dynamics. This behavior persists across the explored range of gas temperatures. Our results indicate that baryonic matter reacts to the global gravitational potential of the virialized halo rather than to the detailed phase-dependent granular structure of FDM. This weak temporal coupling may limit the possibility of directly detecting FDM granularity through local baryonic temporal fluctuations.
\end{abstract}


\maketitle

\section{Introduction}
\label{sec:intro}

Fuzzy Dark Matter (FDM), also known as ultralight bosonic dark matter, describes dark matter as a coherent wave function governed by the Schr\"odinger--Poisson system of equations \cite{Schive:2014dra,Hui:2016,Niemeyer_2020,Hui:2021tkt}. One of the main motivations for this model is that wave effects become relevant at galactic scales, suppressing small-scale structure formation and naturally generating cored density profiles. Numerical simulations of structure formation have shown that virialized halos develop a characteristic core-halo morphology, consisting of a central solitonic core surrounded by an extended fluctuating halo \cite{Schive:2014dra,Mocz:2017wlg,Veltmaat_2018,May_2021,Gotinga2022}. 

The halo region of virialized FDM structures is characterized by a granular density distribution generated by the superposition of  modes of the wave function, and are interpreted as a direct manifestation of the wave nature of dark matter. Their properties have been studied in the context of structure formation simulations \cite{Schive:2014dra,Mocz:2017wlg,Veltmaat_2018,May_2021,Gotinga2022}, local collapse scenarios \cite{periodicas}, and kinetic relaxation processes \cite{Rusos2018,Chen2021}. In particular, recent studies have found that the granular dynamics of FDM halos may induce stochastic effects on the motion of stars, compact objects, and test particles embedded in the halo \cite{CaosFDM,YuZhao2025,ElZant2020,Lancaster_2020}. This motivates the idea that the interference-driven fluctuations of FDM could produce observable imprints on baryonic matter.

At the same time, simulations including baryonic matter have shown that gas dynamically reacts to the gravitational potential sourced by FDM structures \cite{MoczPRL2019,GotingaBarionFDM}. Other cases include the study of the coupled evolution of FDM and an ideal gas in the formation of galaxy-type structures,  \cite{AlvarezGuzman2022}, the stability of Fermion-Boson Stars \cite{AlvarezGuzman2023}, the attractor nature of such stars during the  relaxation process of FDM with gas \cite{FermionBosoStars2024} and potential observable configurations of baryonic matter around FDM vortex solutions \cite{AlvarezTenaGuzman2025}. In particular,  \cite{FermionBosoStars2024} shows that the coupled system can relax toward long-lived core-halo configurations, including scenarios where the baryonic component traces the  structure generated by the FDM distribution. 

However, it is still unclear whether the fast interference-driven fluctuations of core-halo FDM structures produce a direct temporal imprint on the baryonic component. This question is particularly relevant from an observational perspective. If baryonic tracers developed temporal correlations with the granular fluctuations of FDM at galactic scales, such correlations could provide a potential dynamical signature of the wave nature of dark matter. A clear temporal correlation between FDM and ideal gas oscillations has been seen in fundamental oscillation modes of fermion-boson stars \cite{AlvarezGuzman2023}.

Advances in this direction include the results in \cite{Li_2021}, where a mode-expansion is constructed for the FDM, allowing characteristic time-scale oscillations to be associated with energy differences between modes. In \cite{Widmark_2024}, the impact of random-walk behavior and fundamental oscillation modes of the core on test particles is studied. Building on these works, we extend the analysis by coupling an ideal gas to the full FDM dynamics, where the results of \cite{Li_2021}, derived for a pure FDM system, cannot be directly applied. In addition, we include the backreaction of the gas on the FDM dynamics, going beyond the test-particle treatment considered in considered in \cite{Widmark_2024}.

With this motivation, in this work we study the temporal correlations between FDM and baryonic matter, not at core scale, but in the spatially dominant granular structure of galactic size, not in the fundamental low-frequency mode, but in short time-scales, not in spherical symmetry as in \cite{AlvarezGuzman2023}, but in full 3D configurations evolved with the Schr\"odinger-Poisson-Euler system. The idea is to investigate whether observable correlations arise in this regime. For this, we analyze the temporal fluctuations of both components using autocorrelation functions, cross-correlations, and power spectra measured along virtual detectors distributed across a computational domain. We explore different thermodynamic initial conditions for the gas in order to explore the various possible reactions from the gas to small scale FDM fluctuations.

Our analysis reveals a hierarchy of temporal scales in the coupled system. The FDM component exhibits fast, broadband fluctuations associated with interference and granularity, whereas the gas displays smoother and more coherent dynamics dominated by low-frequency modes. More importantly, we find that the temporal cross-correlation between both components remains weak and short-lived. This suggests that the baryonic component modeled with a basic ideal gas does not directly track the fast interference dynamics of FDM, but instead responds primarily to the coarse-grained global gravitational potential of the structure.

The paper is organized as follows. In Section \ref{sec:model} we present the equations of the Schr\"odinger-Poisson-Euler system, the initial conditions of the two components and the methods used for the correlation analysis. In Section \ref{sec:results} we present the results and their analysis. Finally in Section \ref{sec:conclusions} we draw some conclusions.

\section{Physical model and numerical setup}
\label{sec:model}
\subsection{Schr\"odinger Poisson Euler system}
\label{subsec:sep}

We consider a self gravitating two component system composed of a fuzzy dark matter (FDM) component and a baryonic component modeled with an ideal gas (IG). In dimensionless units, the dynamics is governed by the Schr\"odinger-Poisson-Euler (SPE) system \cite{AlvarezGuzman2022, AlvarezGuzman2023, FermionBosoStars2024, AlvarezTenaGuzman2025}

\begin{align}
i\,\partial_t \Psi &= -\frac{1}{2}\nabla^2\Psi + V\Psi,\label{eq:sch_sep}\\
\partial_t \rho + \nabla\cdot(\rho\mathbf{v}) &= 0,\label{eq:cont_sep}\\
\partial_t(\rho\mathbf{v}) + \nabla\cdot\left(\rho\mathbf{v}\otimes\mathbf{v}+ p\,\mathbb{I}\right)&= -\rho\nabla V,\label{eq:mom_sep}\\
\partial_t E +\nabla\cdot\left[(E+p)\mathbf{v}\right] &= -\rho\,\mathbf{v}\cdot\nabla V, \label{eq:energy_sep}\\
\nabla^2 V &= \rho_T - \bar{\rho}_T,\label{eq:pois_sep}
\end{align}

\noindent where the total density is $\rho_T = \rho_{\mathrm{FDM}} + \rho$ with $\rho_{\mathrm{FDM}} = |\Psi|^2$ the mass density of the FDM, $\Psi(\mathbf{x},t)$ describes the macroscopic wave function of the FDM, while $\rho$, $\mathbf{v}$, and $p$ denote the density, velocity field, and pressure of the IG, respectively. The gravitational potential $V$ is sourced by the total density $\rho_T$, with $\bar{\rho}_T$ its spatial average. The system of equations is closed with the ideal gas equation of state,

\begin{equation}
p = (\gamma - 1)\rho e,
\label{eq:eos_ideal}
\end{equation}

\noindent where $e$ is the specific internal energy and $\gamma$ is the adiabatic index. In this case, the total gas energy density in Eq. (\ref{eq:energy_sep}) is given by

\begin{equation}
E = \rho (e + \frac{1}{2} |\mathbf{v}|^2).
\end{equation}

\noindent Thus, equations~\eqref{eq:sch_sep}-\eqref{eq:pois_sep} describe the gravitational coupling between the FDM and IG components. The FDM evolves according to the Schr\"odinger equation, while the IG follows the Euler equations, both coupled through the gravitational potential.

\subsection{Initial conditions and relaxed configurations}
\label{subsec:ic}

The initial conditions for the FDM component are the superposition of various solitonic, randomly located cores, that  undergo a gravitational multimerger. This configuration is motivated by previous studies showing that multimergers of solitonic configurations provide  the formation of virialized structures in ultralight bosonic dark matter \cite{Schive:2014hza}. 
In particular, numerical simulations show that the collision and relaxation of solitonic cores lead to the formation of stable core-halo configurations \cite{Schwabe:2016}, and generate turbulent, granular halos exhibiting universal structural properties \cite{mocz19}. Also, previous work has shown that such configurations form under different numerical setups and boundary conditions, supporting their physical robustness \cite{periodicas}. Taken together, these results justify the use of multimerger initial conditions as a controlled and physically motivated approach to generate virialized FDM core-halo systems.

On the other hand, the gas component is initialized from a random Gaussian field constructed in Fourier space. Specifically, an auxiliary field is defined as

\begin{equation}
\Psi_{\mathrm{aux}}(\mathbf{x}) = A\,\mathcal{F}^{-1}\left\{ e^{-|\mathbf{p}|^2}\,e^{i\phi(\mathbf{p})} \right\},\label{eq:gaussianp}
\end{equation}

\noindent where $\mathcal{F}^{-1}$ denotes the inverse Fourier transform, $\mathbf{p}$ are momentum-space coordinates, and $\phi(\mathbf{p})$ is a random phase uniformly distributed in $[0,2\pi)$ for each mode. The normalization constant $A$ is chosen such that the total IG mass satisfies the desired value. The initial gas density is then defined as

\begin{equation}
\rho(\mathbf{x},0) = |\Psi_{\mathrm{aux}}(\mathbf{x})|^2.
\end{equation}

\noindent This method generates a smooth random field without introducing preferred directions or scales beyond those imposed by the spectral filter, in our case the Gaussian filter $e^{-|\mathbf{p}|^2}$, which suppresses high-wavenumber modes and therefore determines the characteristic scale of the fluctuations. Similar initializations have been employed in previous studies of coupled bosonic dark matter and baryonic fluids, where random density perturbations are used to study the dynamical response of the gas to the underlying FDM structure \cite{AlvarezTenaGuzman2025,FermionBosoStars2024}.

The initial gas velocity is set to zero

\begin{equation}
\mathbf{v}(\mathbf{x},0)=0,
\end{equation}

\noindent whereas the initial pressure is prescribed through the polytropic equation of state

\begin{equation}
p(\mathbf{x},0)=K\,\rho(\mathbf{x},0)^{\gamma},
\end{equation}

\noindent which is used only to define the initial thermodynamic state and not during the evolution. For $t>0$, the IG evolves according to the equation of state \eqref{eq:eos_ideal}.

With these initial conditions, the multimerger of FDM cores and the evolution of the gas fluctuations evolve toward a virialized configuration. Since our system uses random initial conditions and follows the self-consistent gravitational collapse and relaxation of both the FDM and the gas, and moreover the final relaxed state emerges naturally from chaotic fluctuations, it represents a generic and statistically valid configuration without the need for an ensemble of runs using a smoothed idealized set initial conditions.

\subsection{Diagnostics}
\label{subsec:diagnostics}

Once the FDM multi-core merger coupled to the ideal gas virializes, we start the search of a potential correlation between the two components, which consists of various steps described below.

{\it Normalized density fluctuation.} We characterize the fluctuations in time of the FDM and the IG   around their time-averaged profiles. For each species we introduce the temporal normalized density fluctuation

\begin{align}
\Delta_{\mathrm{FDM}} &\equiv
  \frac{\rho_{\mathrm{FDM}}        - \langle\rho_{\mathrm{FDM}}\rangle_t}   {\langle\rho_{\mathrm{FDM}}\rangle_t},\label{eq:delta_fdm}\\
\Delta_{\mathrm{IG}} &\equiv  \frac{\rho   - \langle\rho\rangle_t}  {\langle\rho\rangle_t}, \label{eq:delta_ig}
\end{align}

\noindent where $\langle\cdot\rangle_t$ denotes a temporal average on a given time window.

{\it Detectors.} We do not process the time evolution of densities in the whole numerical domain, we analyze the evolution along $n$ arbitrarily chosen lines parallel to the $z$-axis that we call {\it detector lines}. Each detector line $j$ ($j = 1,\dots,n$) is at a fixed transverse position $(x_j,y_j)$, defining the sampling line $\{(x_j,y_j,z)\}$. If $N_z$ is the number of points of the numerical domain along the $z-$direction, then we define $N_z$ {\it detector points} along each detector line. At each detector point, of each detector line, we record the time series $\Delta_{\mathrm{FDM}}^{(j)}(z,t)$ and $\Delta_{\mathrm{IG}}^{(j)}(z,t)$. In this way, we have an ensemble of detector points that provides a statistical sampling of different regions of the system as the core-halo structure evolves. In what follows,  the detector line label $j$ is omitted when expressions apply to any detector.

{\it Correlation functions.} For real-valued signals $f(z,t)$ and $g(z,t)$ sampled along a given detector line, the temporal cross-correlation with lag $\tau$ is defined as \cite{Pope2000}

\begin{equation}
C_{fg}(z,\tau) \equiv
  \bigl\langle f(z,t)\,g(z,t+\tau)\bigr\rangle_t,\label{eq:corr_def}
\end{equation}

\noindent the autocorrelation functions are defined as

\begin{align}
C_{\mathrm{FDM}}(z,\tau) &\equiv  \bigl\langle    \Delta_{\mathrm{FDM}}(z,t)\,\Delta_{\mathrm{FDM}}(z,t+\tau)  \bigr\rangle_t,
\label{eq:auto_fdm}\\
C_{\mathrm{IG}}(z,\tau) &\equiv  \bigl\langle    \Delta_{\mathrm{IG}}(z,t)\,\Delta_{\mathrm{IG}}(z,t+\tau)  \bigr\rangle_t,
\label{eq:auto_ig}
\end{align}

\noindent and the cross-correlation between both components is given by

\begin{equation}
C_{\times}(z,\tau) \equiv  \bigl\langle    \Delta_{\mathrm{FDM}}(z,t)\,\Delta_{\mathrm{IG}}(z,t+\tau)  \bigr\rangle_t.
\label{eq:cross_corr}
\end{equation}

\noindent In particular, $C_{\alpha}(z,0) = \langle\Delta_\alpha^2\rangle_t$ corresponds to the local temporal variance of the  species $\alpha$.

{\it Power spectra and cross-spectrum.} The spectral content of the fluctuations is obtained from the temporal Fourier transform

\begin{equation}
\hat{\Delta}(z,f) \equiv  \int_{-\infty}^{\infty} \Delta(z,t)\,e^{-i2\pi f t}\,dt,
\label{eq:fourier_def}
\end{equation}

\noindent the power spectral densities are defined as

\begin{align}
P_{\mathrm{FDM}}(z,f) &\equiv  \bigl|\hat{\Delta}_{\mathrm{FDM}}(z,f)\bigr|^2,\label{eq:ps_fdm}\\
P_{\mathrm{IG}}(z,f) &\equiv  \bigl|\hat{\Delta}_{\mathrm{IG}}(z,f)\bigr|^2,\label{eq:ps_ig}
\end{align}

\noindent whereas the cross-power spectrum is defined as

\begin{equation}
P_{\times}(z,f) \equiv  \hat{\Delta}_{\mathrm{FDM}}(z,f)\,  \hat{\Delta}_{\mathrm{IG}}^{\,*}(z,f),
\label{eq:cross_spec}
\end{equation}

\noindent where $*$ denotes complex conjugation. By the Wiener-Khinchin theorem \cite{Papoulis2002}, $P_\alpha(z,f)$ is the Fourier transform of $C_\alpha(z,\tau)$, and $P_{\times}(z,f)$ is the Fourier transform of $C_{\times}(z,\tau)$.

{\it Characteristic time scales.} We define the normalized autocorrelation function as

\begin{equation}
\tilde{C}_\alpha(z,\tau) \equiv \frac{C_\alpha(z,\tau)}{C_\alpha(z,0)},
\label{eq:norm_acf}
\end{equation}

\noindent so that $\tilde{C}_\alpha(z,0)=1$ by construction. The \emph{integral correlation time} is defined as \cite{Pope2000}

\begin{equation}
\tau_{\alpha}(z) \equiv \int_{0}^{\tau_0} \tilde{C}_\alpha(z,\tau)\,d\tau,
\label{eq:tau_int}
\end{equation}

\noindent where $\tau_0$ denotes the first zero crossing of $C_\alpha(z,\tau)$. This definition isolates the primary positive correlation lobe and avoids cancellations arising from oscillatory behavior at larger time lags.

For the cross-correlation, we define the normalized function as

\begin{equation}
\tilde{C}_{\times}(z,\tau) \equiv \frac{C_{\times}(z,\tau)}{\max_{\tau} |C_{\times}(z,\tau)|},
\label{eq:norm_ccf}
\end{equation}

\noindent where the normalization is taken with respect to the maximum amplitude of $|C_{\times}|$, which in general does not occur at $\tau=0$. The corresponding integrated correlation time is defined as \cite{Pope2000}

\begin{equation}
\tau_{\times}(z) \equiv \int_{\tau_{\mathrm{lag}}}^{\tau_0^{\times}}|\tilde{C}_{\times}(z,\tau)|\,d\tau,
\label{eq:tau_int_cross}
\end{equation}

\noindent where $\tau_{\mathrm{lag}}$ is the lag at which $|C_{\times}(z,\tau)|$ approaches its maximum, and $\tau_0^{\times}$ denotes the first zero crossing of $C_{\times}(z,\tau)$ for $\tau>\tau_{\mathrm{lag}}$. The use of the absolute value avoids cancellations due to sign changes in the cross-correlation function.

We use all  these elements to characterize the temporal coherence of each component and of their dynamical coupling.

\subsection{Parameter space and numerical set up}
\label{subsec:numerics}

The system is evolved using the code \texttt{CAFE-FDM}, introduced in \cite{AlvarezGuzman2022}. The time integration is performed with a fourth-order Runge-Kutta (RK4) scheme. The right-hand side of the Schr\"odinger equation is discretized using fourth-order finite-difference stencils, while the hydrodynamic equations are solved using a finite-volume approach based on high-resolution shock-capturing (HRSC) methods. Specifically, we use the \texttt{minmod} limiter for the reconstruction of primitive variables and the approximate HLLE flux formula to approximate numerical fluxes. Periodic boundary conditions are imposed in all spatial directions. The gravitational potential is calculated by solving the Poisson equation using an FFT-based solver, which is appropriate for periodic domains.

All simulations are carried out in a cubic domain $[-40,40]^3$, evolved up to a final time $t = 200$. The spatial resolution is $N = 128$ grid points per dimension, corresponding to a uniform grid spacing of $h = 0.625$. The time step is chosen as $\Delta t = 0.064\,h^2 = 0.025$, ensuring stability of the evolution. 

The FDM component is initialized as a multimerger configuration consisting of 15 solitonic cores distributed randomly within the computational domain. Each core is initially at rest, with central density in the range $[0.8,\,1.2]$. This setup leads to a sequence of mergers that drives the system toward a virialized core-halo structure.

For the gas component, we consider a monoatomic ideal gas with adiabatic index $\gamma = 5/3$. The initial Gaussian density field (\ref{eq:gaussianp}) is normalized such that the total gas mass corresponds to $10\%$ of the total FDM mass, since we consider the scenario dominated by FDM. The initial thermodynamic state is specified through a polytropic constant, and we explore the values $K = 10^{-3},\,10^{-2},\,10^{-1},\,1$, in order to see the impact of the initial gas temperature on the dynamical coupling. Smaller values of $K$ correspond to colder initial conditions, while larger values represent progressively hotter gas distributions.

\section{Results}
\label{sec:results}
\subsection{Dynamical evolution and virialization}
\label{subsec:K1_evolution}

\begin{figure}
\centering
\includegraphics[width=8cm]{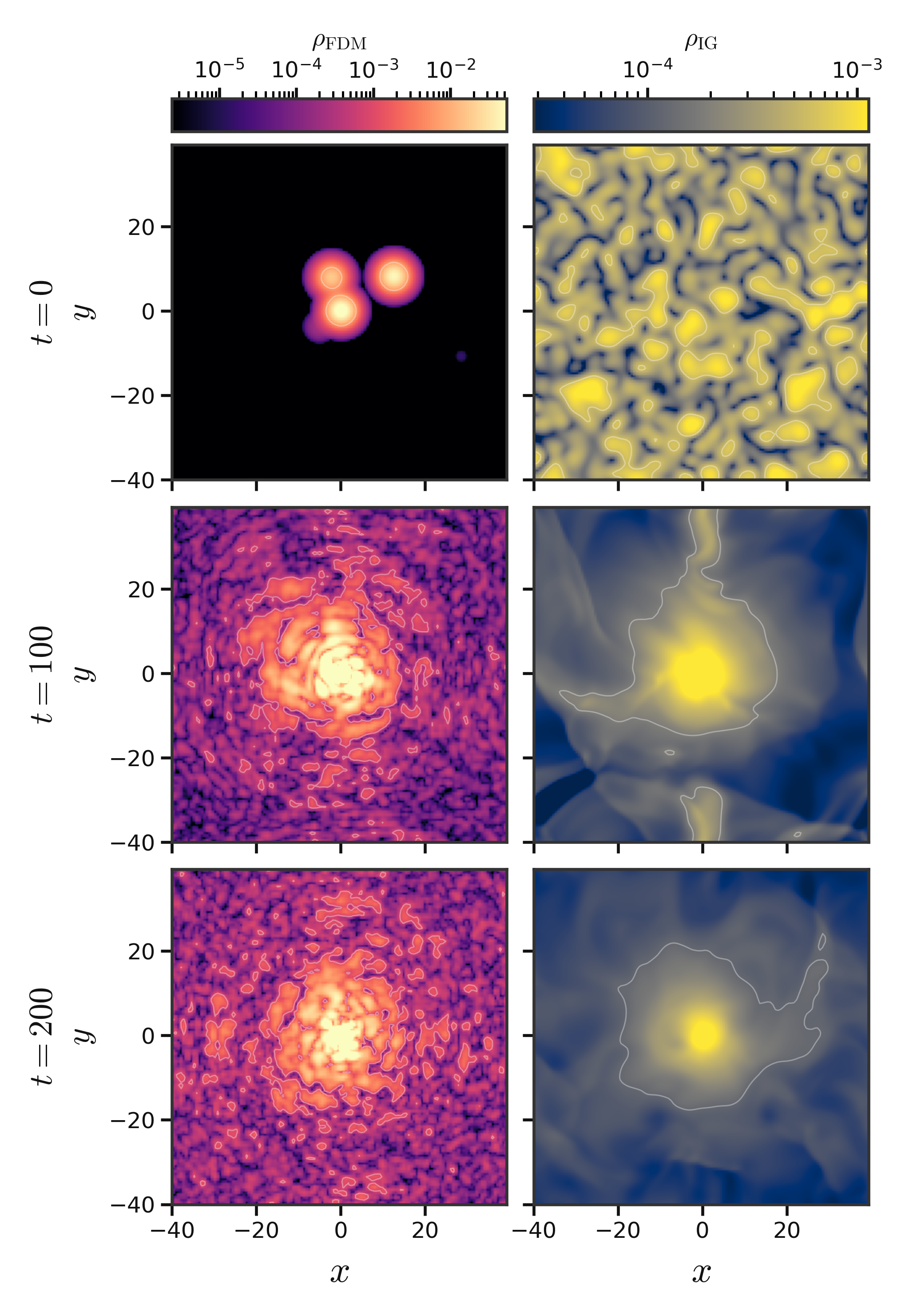}
\caption{Density of FDM and IG in the plane $z = 0$ for the case $K=1$ at $t=0$ (top), $t=100$ (middle), and $t=200$ (bottom). Left panels show the FDM density, while right panels correspond to the IG density. The initial configuration consists of multiple isolated solitonic cores for the FDM and a smooth random field for the gas. As the system evolves, gravitational multimergers lead to the formation of a central core-halo structure. The FDM exhibits granular interference patterns, whereas the gas develops a smoother, pressure-supported distribution responding seemingly to the large-scale potential.}
\label{fig:slices_K1}
\end{figure}

As a representative example, we present the analysis of the system for the case $K=1$. Figure~\ref{fig:slices_K1} shows density slices in the plane $z= 0$ for both the FDM and gas components at $t=0$, $t=100$, and $t=200$. 

At $t=0$ we show the initial conditions. As the system evolves, gravitational interactions drive a sequence of mergers among the solitons, leading to the formation of a central overdensity surrounded by a fluctuating halo. The FDM exhibits the characteristic granularity, whereas the gas develops a smoother and more extended distribution, driven by gas pressure and the large scale gravitational potential.

By $t=100$, the system has already formed a well-defined core-halo structure. At later times ($t=200$), the configuration remains qualitatively stable. The FDM remains with a compact solitonic core embedded in a granular halo, while the gas forms a smoother component tracing the large-scale potential.

\begin{figure}
\centering
\includegraphics[width=8cm]{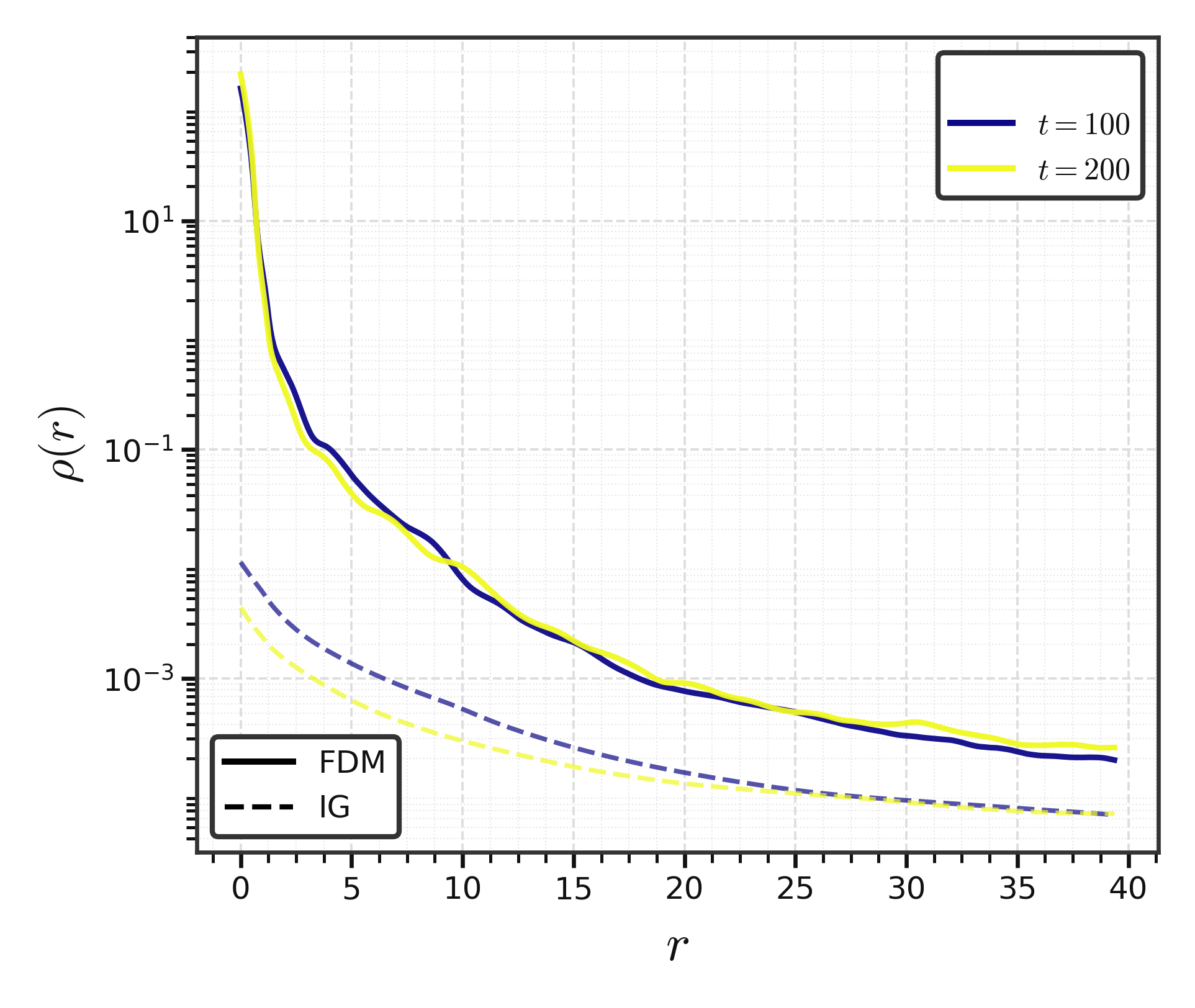}
\caption{Spherically averaged density profiles for the FDM (solid lines) and gas (dashed lines) components at $t=100$ and $t=200$ for $K=1$. Both components exhibit stable radial distributions at late times, indicating that the system has reached a quasi-stationary configuration. The FDM profile shows a central core surrounded by an extended halo, while the gas remains smoother and subdominant. The persistence of the central structure is consistent with the presence of a long-lived core configuration.}
\label{fig:radial_profiles_K1}
\end{figure}

This behavior is further quantified in Fig.~\ref{fig:radial_profiles_K1}, where we show the solid-angle averaged density profiles at $t=100$ and $t=200$. Both components exhibit stable radial profiles, with the FDM dominated by a central core and an extended halo, while the gas remains subdominant and smoother. The persistence of the central structure is consistent with the interpretation that the core region corresponds to a long-lived attractor solution of the coupled system, namely a fermion-boson star configuration \cite{FermionBosoStars2024}.

\begin{figure}
\centering
\includegraphics[width=8cm]{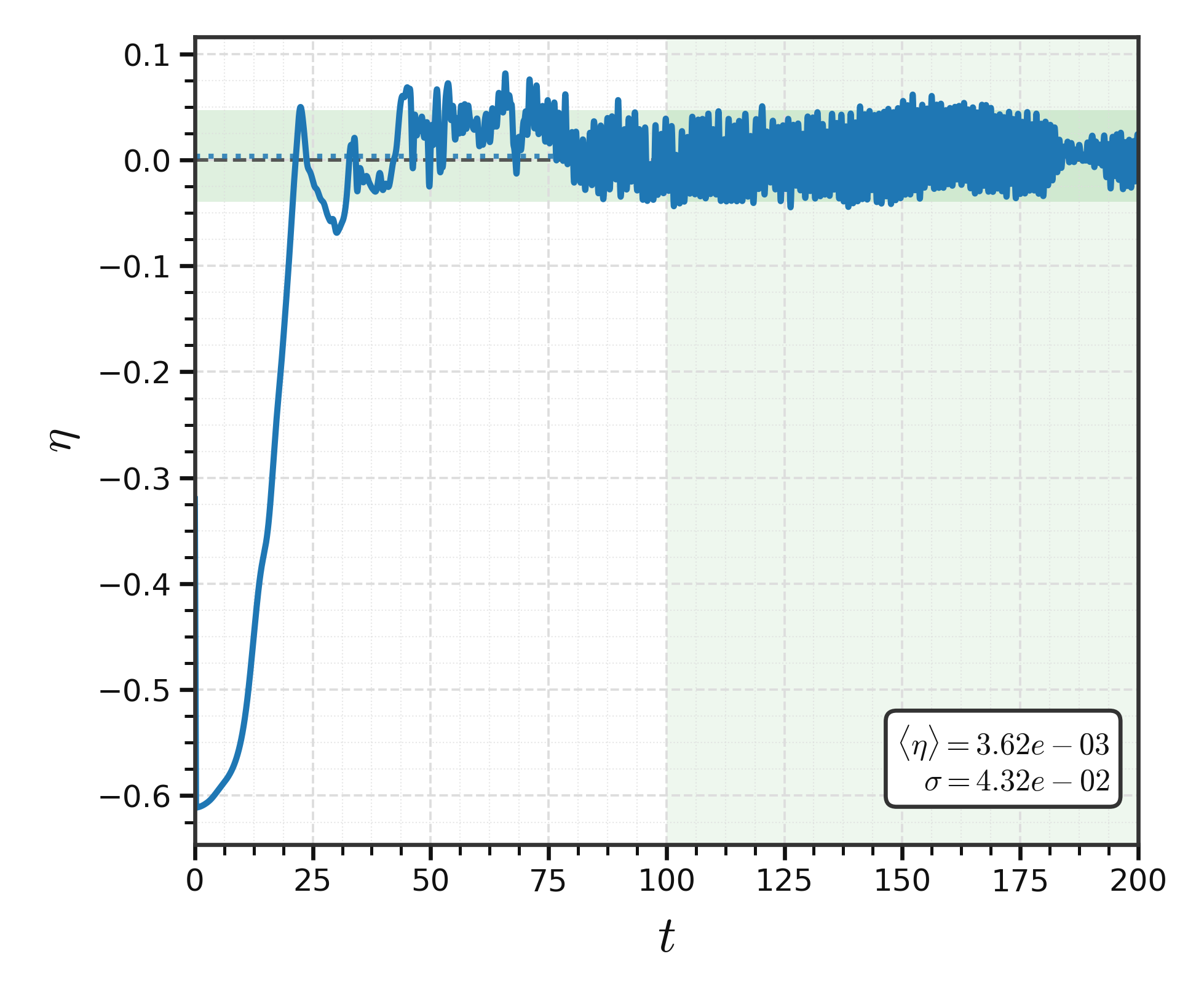}
\caption{Time evolution of the virial quantity $\eta$ for the case $K=1$. After an initial transient associated with the multimerger dynamics, the system approaches $\eta \simeq 0$ and remains close to this value for $t \gtrsim 100$, indicating that virial equilibrium has been achieved. The shaded region highlights the time interval used for the statistical analysis.}
\label{fig:virial_K1}
\end{figure}

Finally, in order to determine when the structure virializes, in Fig.~\ref{fig:virial_K1} we show the evolution of the virial quantity $\eta(t) = \frac{2K + W + 3U}{|W|}$, where $K$, $W$, and $U$ denote the total kinetic, gravitational, and internal energies, respectively. After an initial transient dominated by mergers, the system approaches $\eta \simeq 0$ and remains close to this value for $t \gtrsim 100$, indicating that the configuration evolves around virial equilibrium.

Based on this behavior,  the temporal statistical averages presented in what follows are calculated over the interval $t\in[100,200]$. In this time window, the virial parameter fluctuates around the mean $\langle\eta\rangle_t\simeq0$, indicating that the system has reached a virialized and approximately statistically stationary state. As an illustrative exaple, n physical units, for a boson mass $m_B=10^{-22}$eV and a box of side $L=20$kpc, the total FDM mass is $10^{10}{\rm M}_{\odot}$ and the analysis time-window $100$ to $200$ dimensionless units, lasts  $\sim 0.32$Gyr.

\subsection{Temporal fluctuations and correlation analysis}
\label{subsec:correlations_K1}

{\it Normalized density fluctuations.} In order to characterize the temporal fluctuations of the virialized system, we introduce a set of $n=16$ detector lines distributed as a $4\times4$ grid, at positions $(x_i,y_j) \in \{-30,-10,10,30\} \times \{-30,-10,10,30\}$, within the $[-40,40]^3$ numerical domain. As described above, each detector records the properties of the system along a line parallel to the $z$-axis, allowing us to record time series of the density fluctuations at 128 positions per detector line.

At each detector point, we analyze the normalized density fluctuations
$\Delta_{\mathrm{FDM}}(z,t)$ and $\Delta_{\mathrm{IG}}(z,t)$, which measure the temporal deviations of each component with respect to their time-averaged profiles. These quantities provide a local characterization near each detector line of the dynamical activity of the system.

\begin{figure}
\centering
\includegraphics[width=8cm]{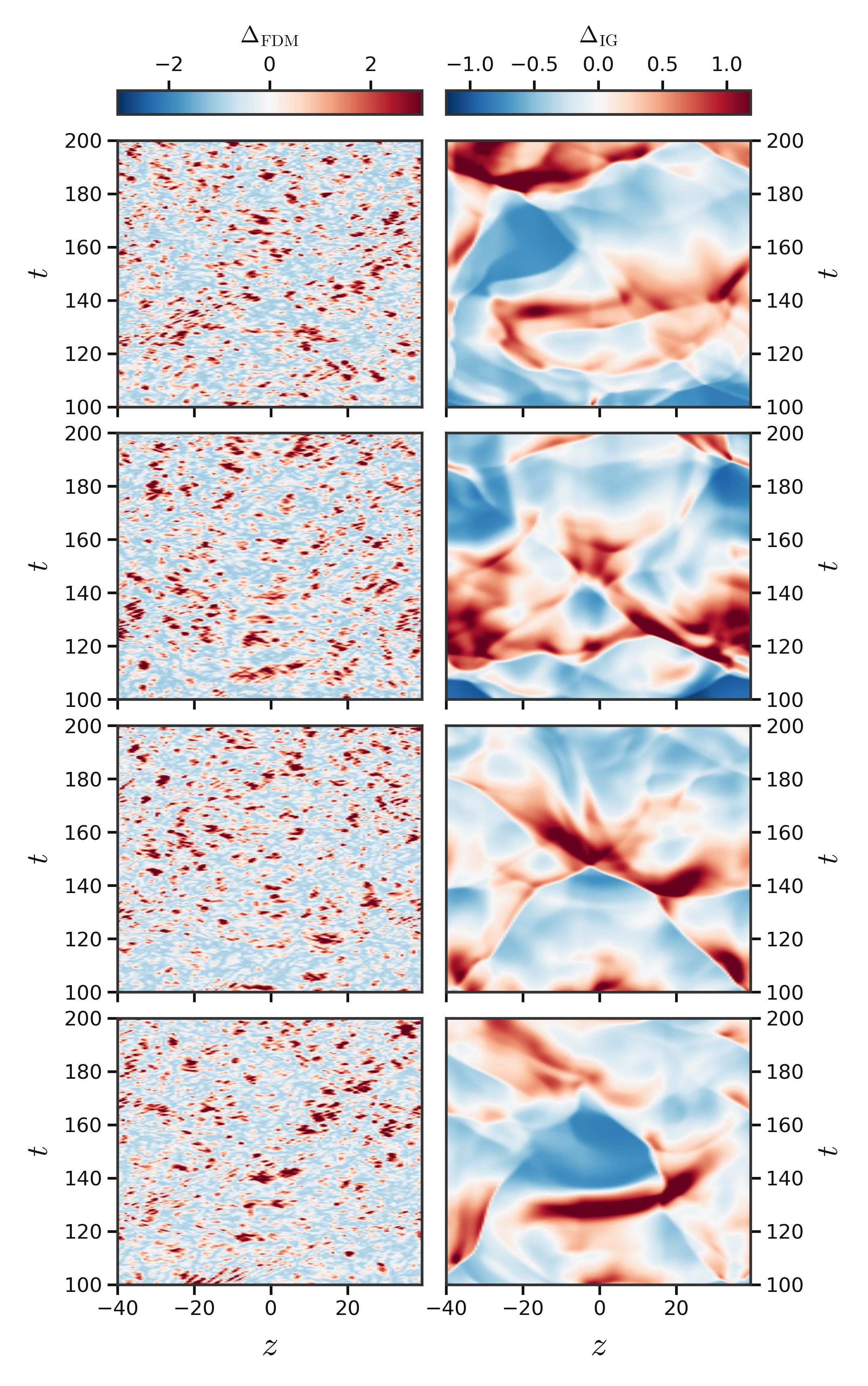}
\caption{Normalized density fluctuations $\Delta(z,t)$ measured at four representative detector lines for the case $K=1$. 
The horizontal axis represents the location of the 128 detector points along the $z-$axis for each of these four detector lines, whereas the vertical axis represents $t$. Each row corresponds to a different detector line, while columns show the FDM (left) and gas (right) components. The FDM fluctuations display rapid temporal variations and small-scale structures associated with wave interference and granular dynamics. In contrast, the gas component exhibits smoother, large-scale fluctuations, reflecting the effect of pressure support. These differences anticipate the different temporal correlation properties of both components.}
\label{fig:delta_K1_detectors}
\end{figure}

Figure~\ref{fig:delta_K1_detectors} shows representative examples of these fluctuations for four of the sixteen detector lines. The left/right panels correspond to the FDM/IG component. The FDM fluctuations exhibit rapidly varying, small-scale structures, reflecting the interference-driven granularity of the wave dynamics. In contrast, the gas fluctuations are significantly smoother and display larger-scale coherent features, consistent with the pressure support of the gas.

This qualitative difference highlights the different dynamical regimes of the two components, that is, while the FDM evolves through fast  interference-dominated processes, the gas responds on longer time scales, smoothing out small-scale variations. These properties are expected to manifest in their respective temporal correlations.

\begin{figure}
\centering
\includegraphics[width=8cm]{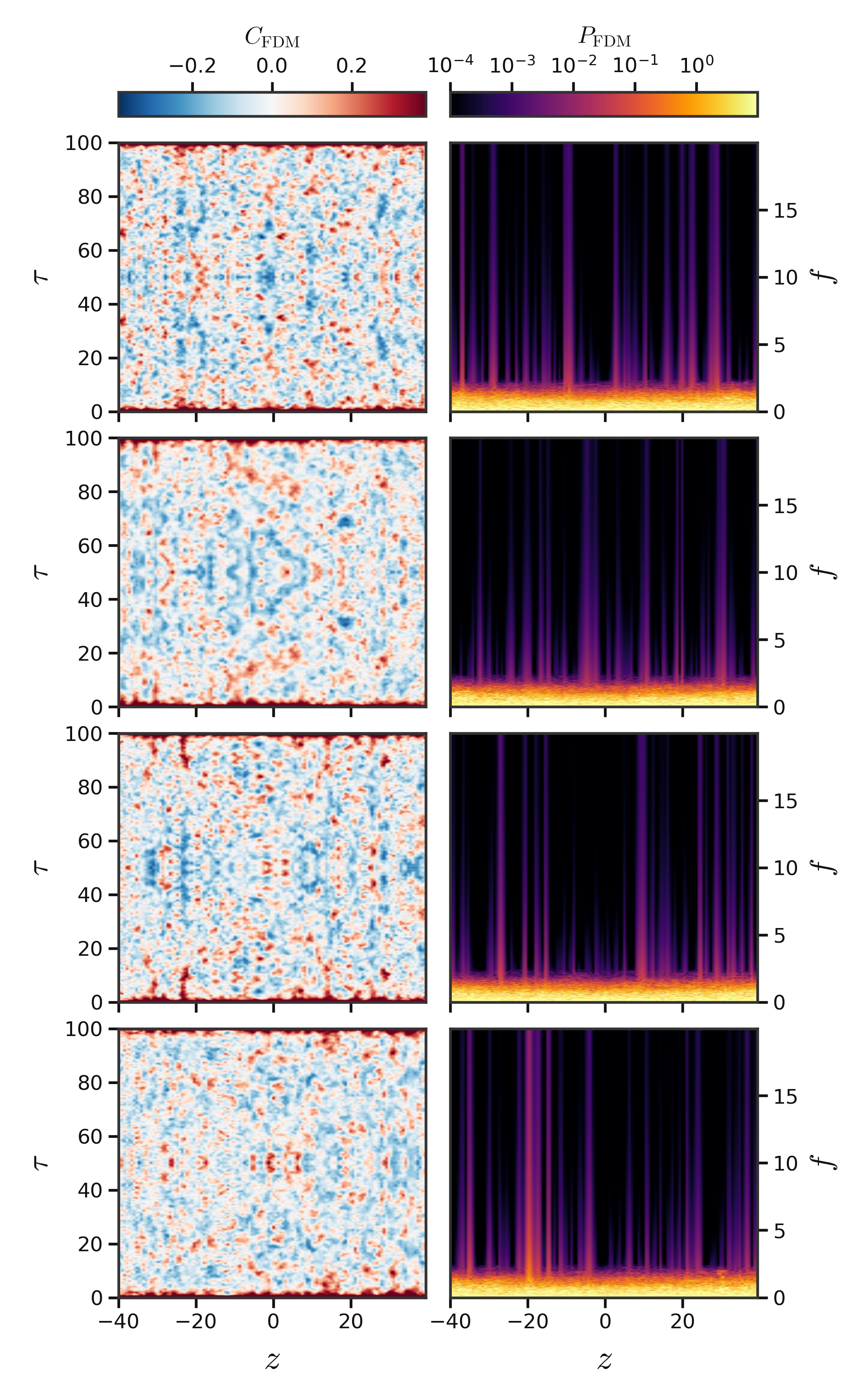}
\caption{Temporal autocorrelation functions $C_{\mathrm{FDM}}(z,\tau)$ (left) and power spectra $P_{\mathrm{FDM}}(z,f)$ (right) for four representative detector lines in the case $K=1$. The horizontal axis corresponds to the 128 detector points along the $z-$direction of each detector line; the vertical axis of left the panel is $\tau$, and the vertical axis of the right panel is frequency. The FDM component exhibits rapidly varying correlations and short coherence times. The corresponding spectra show a broad distribution of frequencies with significant high-frequency content, reflecting the granular and interference-driven dynamics of the FDM halo.}
\label{fig:corr_fdm_K1}
\end{figure}

\begin{figure}
\centering
\includegraphics[width=\linewidth]{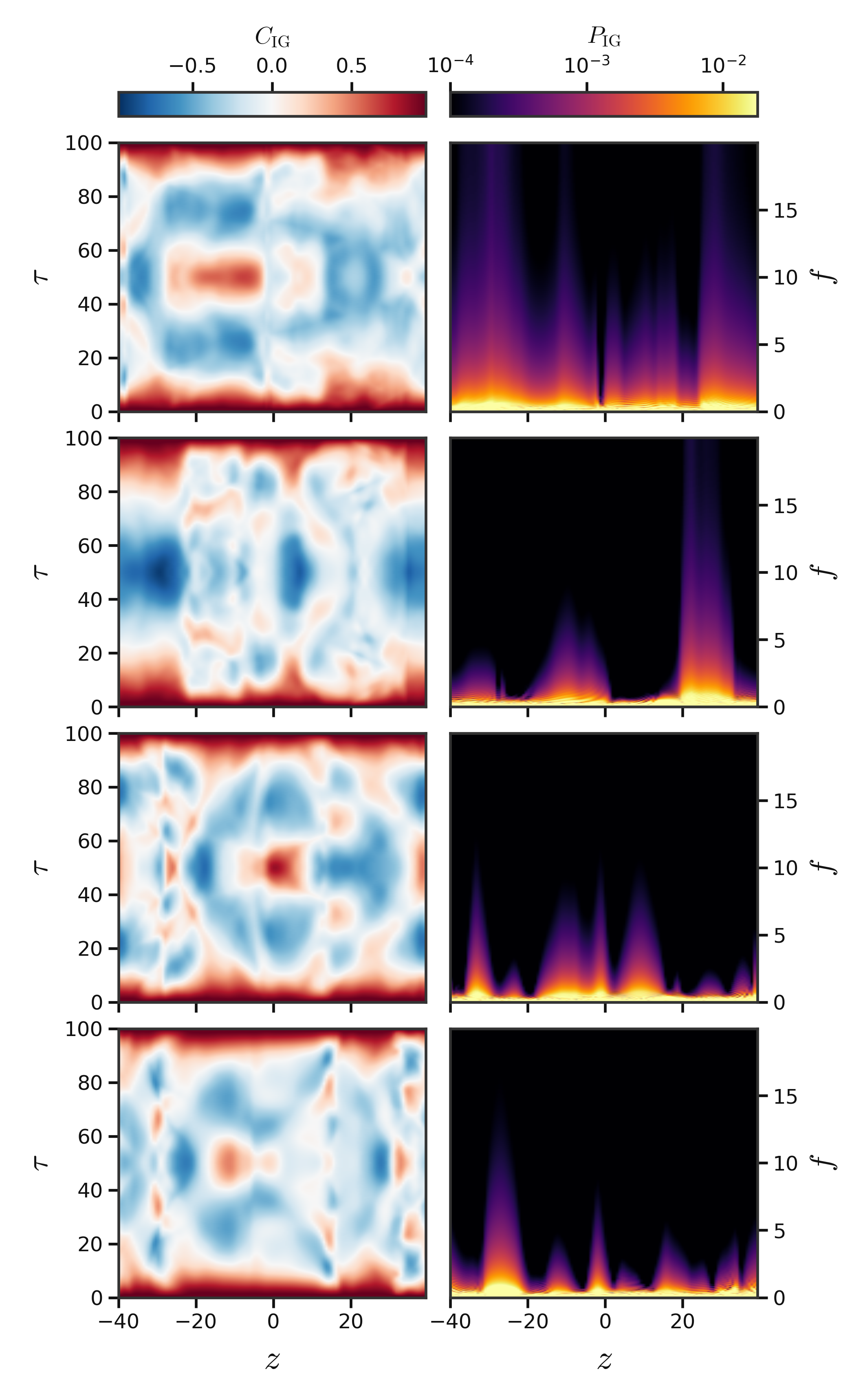}
\caption{Temporal autocorrelation functions $C_{\mathrm{IG}}(z,\tau)$ (left) and power spectra $P_{\mathrm{IG}}(z,f)$ (right) for the same detector lines as in Fig.~\ref{fig:corr_fdm_K1}. The horizontal axis corresponds to the 128 detector points along the $z-$direction of each detector line; the vertical axis of left panel is  $\tau$, and the vertical axis of the right panel is frequency. In contrast to the FDM, the gas component exhibits smoother correlations with longer temporal coherence. The spectra are dominated by low-frequency modes, indicating slower, large-scale dynamics governed by pressure support.}
\label{fig:corr_ig_K1}
\end{figure}

{\it Autocorrelations.} We now quantify the temporal structure of the fluctuations through their autocorrelation and cross-correlation functions, together with their associated power spectra. Figures~\ref{fig:corr_fdm_K1} and \ref{fig:corr_ig_K1} show the temporal
autocorrelation functions $C_{\mathrm{FDM}}(z,\tau)$ and $C_{\mathrm{IG}}(z,\tau)$, along with their corresponding power spectra,
for four representative detectors.

The FDM component exhibits rapidly varying correlations with short coherence in time. The autocorrelation decays quickly with
increasing lag, reflecting the highly dynamical, interference-driven nature of the wave field. Consistently, the power spectra display a
broad distribution of frequencies with significant high-frequency content, indicating the presence of fast, small-scale temporal
fluctuations.

In contrast, the gas component shows significantly smoother autocorrelation patterns, extending over larger values of $\tau$.
This behavior indicates longer memory and slower temporal evolution. The corresponding spectra are dominated by low-frequency modes,
revealing that the gas dynamics is governed by large-scale, pressure-supported structures, very different from the high-frequency behavior of the FDM.

\begin{figure}
\centering
\includegraphics[width=8cm]{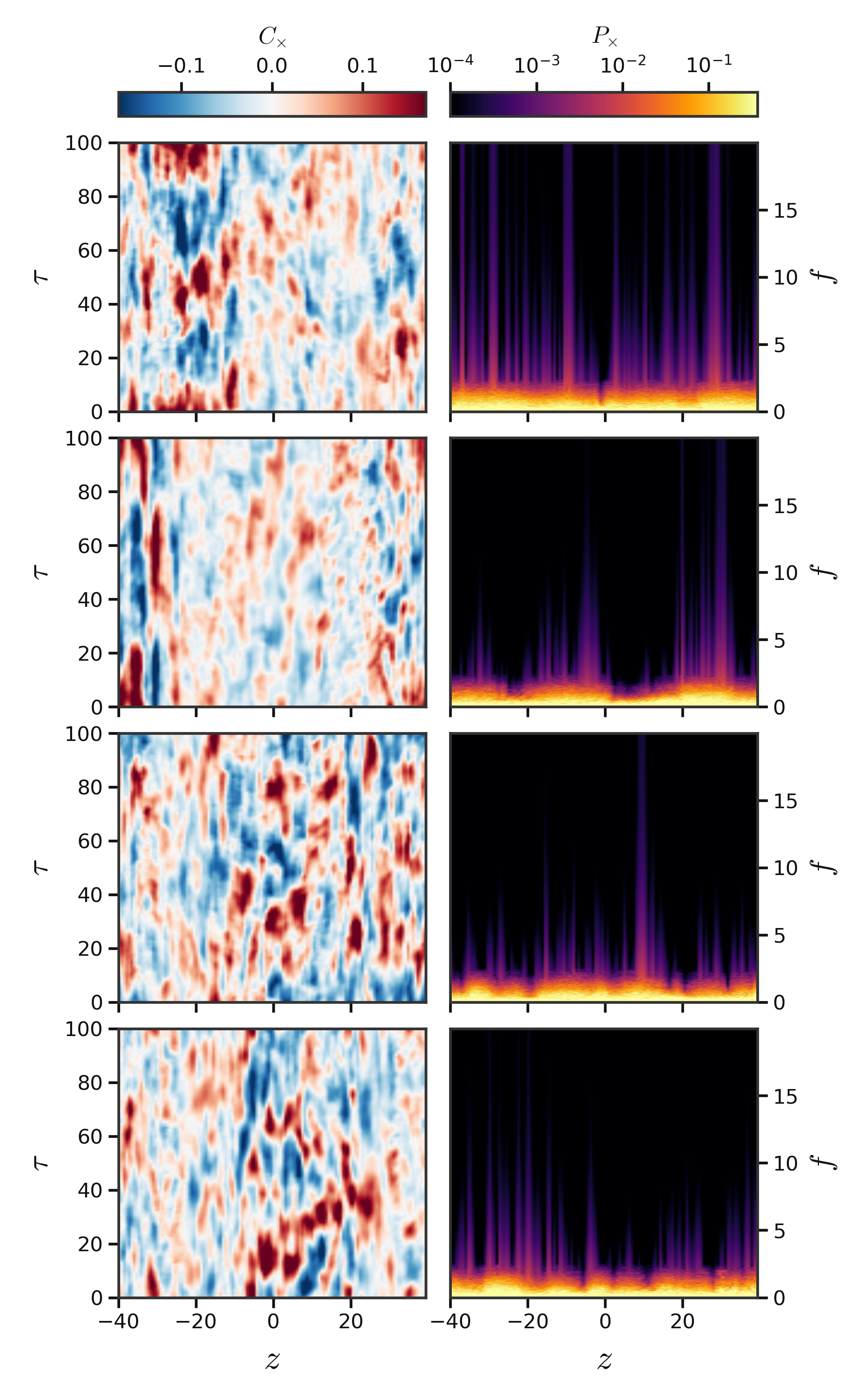}
\caption{Temporal cross-correlation functions $C_{\times}(z,\tau)$ (left) and cross-power spectra $P_{\times}(z,f)$ (right) for four representative detector lines, for the case $K=1$. The horizontal axis corresponds to the 128 detector points along the $z-$direction of each detector line; the vertical axis of left panel is  $\tau$, and the vertical axis of the right panel is frequency. The cross-correlation exhibits intermediate behavior between the FDM and gas components, with coherent structures but reduced persistence. The cross-spectrum is dominated by low-frequency modes, as seen in the right panles, indicating that the coupling between both components is primarily driven by large-scale, slowly varying gravitational interactions.}
\label{fig:corr_cross_K1}
\end{figure}

{\it Cross-correlation.} The temporal coupling between the two components is illustrated in Fig.~\ref{fig:corr_cross_K1}, which shows the cross-correlation function $C_{\times}(z,\tau)$ and the corresponding cross-spectrum $P_{\times}(z,f)$. The cross-correlation exhibits intermediate behavior between the two components, it displays coherent structures in both $z$ and $\tau$, but with reduced amplitude and less persistence compared to the gas autocorrelation. This indicates that, while the gas responds to the large spatial scale of the 
gravitational potential generated by the FDM matter distribution, the coupling is not instantaneous nor fully coherent at all scales.

The cross-spectrum further reveals that the coupling is dominated by low-frequency modes, as can be seen in the panels at the right, in which the cross-spectrum reaches its maximum at low frequencies, with only limited contribution from high-frequency fluctuations. This suggests that the interaction between the FDM and gas components is mediated primarily by the large-scale, slowly varying gravitational potential, while the fast interference-driven dynamics of the FDM remains only weakly imprinted on the gas.

Taken together, these results demonstrate a clear separation of temporal scales in the system, the FDM is characterized by fast, broadband
fluctuations, the gas by slow, coherent dynamics, and their coupling by an intermediate regime dominated by low-frequency interactions. This hierarchy of time scales is a key ingredient to understand the dynamical interaction between both components.

{\it Integrated correlation time.} The qualitative differences observed in the autocorrelation functions and power spectra can be quantified through the integrated correlation time defined in Eq. (\ref{eq:tau_int}), which provides a compact measure of the temporal coherence of the fluctuations.

\begin{figure}
\centering
\includegraphics[width=8cm]{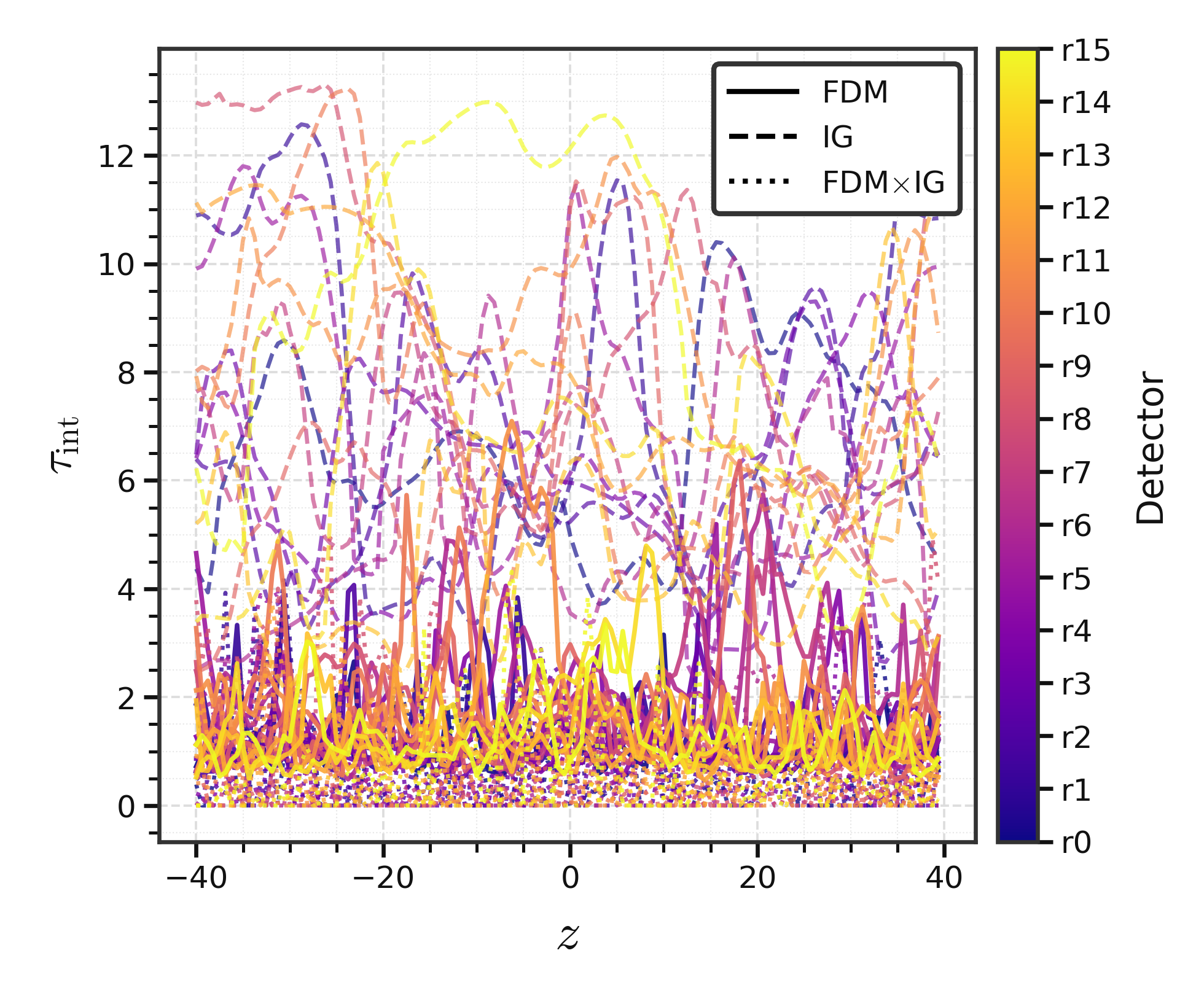}
\caption{Integrated correlation time $\tau_{\mathrm{FDM}}(z)$ (solid), $\tau_{\mathrm{IG}}(z)$ (dashed), and $\tau_{\times}(z)$ (dotted) for all 16 detector lines for the case $K=1$. The FDM exhibits short correlation times, the gas significantly longer ones, and the cross-correlation is generally smaller than the FDM, indicating rapidly decorrelating FDM-gas coupling with noticeable spatial variations.}
\label{fig:tau_int_K1}
\end{figure}

Figure~\ref{fig:tau_int_K1} shows $\tau_{\mathrm{FDM}}(z)$, $\tau_{\mathrm{IG}}(z)$, and $\tau_{\times}(z)$ for all detectors. The FDM component exhibits short correlation times, $\tau_{\mathrm{FDM}} \sim \mathcal{O}(1)$, while the gas reaches significantly larger values, $\tau_{\mathrm{IG}} \sim \mathcal{O}(10)$, consistent with its smoother, pressure-supported dynamics.

The cross-correlation time $\tau_{\times}(z)$ is typically smaller than $\tau_{\mathrm{FDM}}(z)$, although local variations are present due to spatial inhomogeneities and residual oscillatory behavior. This indicates that the mutual correlation between both components decorrelates on time scales shorter than those characterizing the intrinsic FDM fluctuations.

Physically, this implies that the baryonic component does not follow the rapid, interference-driven dynamics of the FDM at the level of temporal fluctuations. Instead, the coupling between both components is limited to short temporal intervals, with the gas responding primarily to the coarse-grained, rather than locally fluctuating, gravitational potential.

The relative ordering of the time scales is not strictly preserved at every position along $z$, reflecting the complex and inhomogeneous structure of the virialized configuration. Nevertheless, the overall behavior suggests the existence of a well-defined hierarchy of temporal scales, which becomes more evident when considering averages over the ensemble of detector points.

\begin{figure}
\centering
\includegraphics[width=8cm]{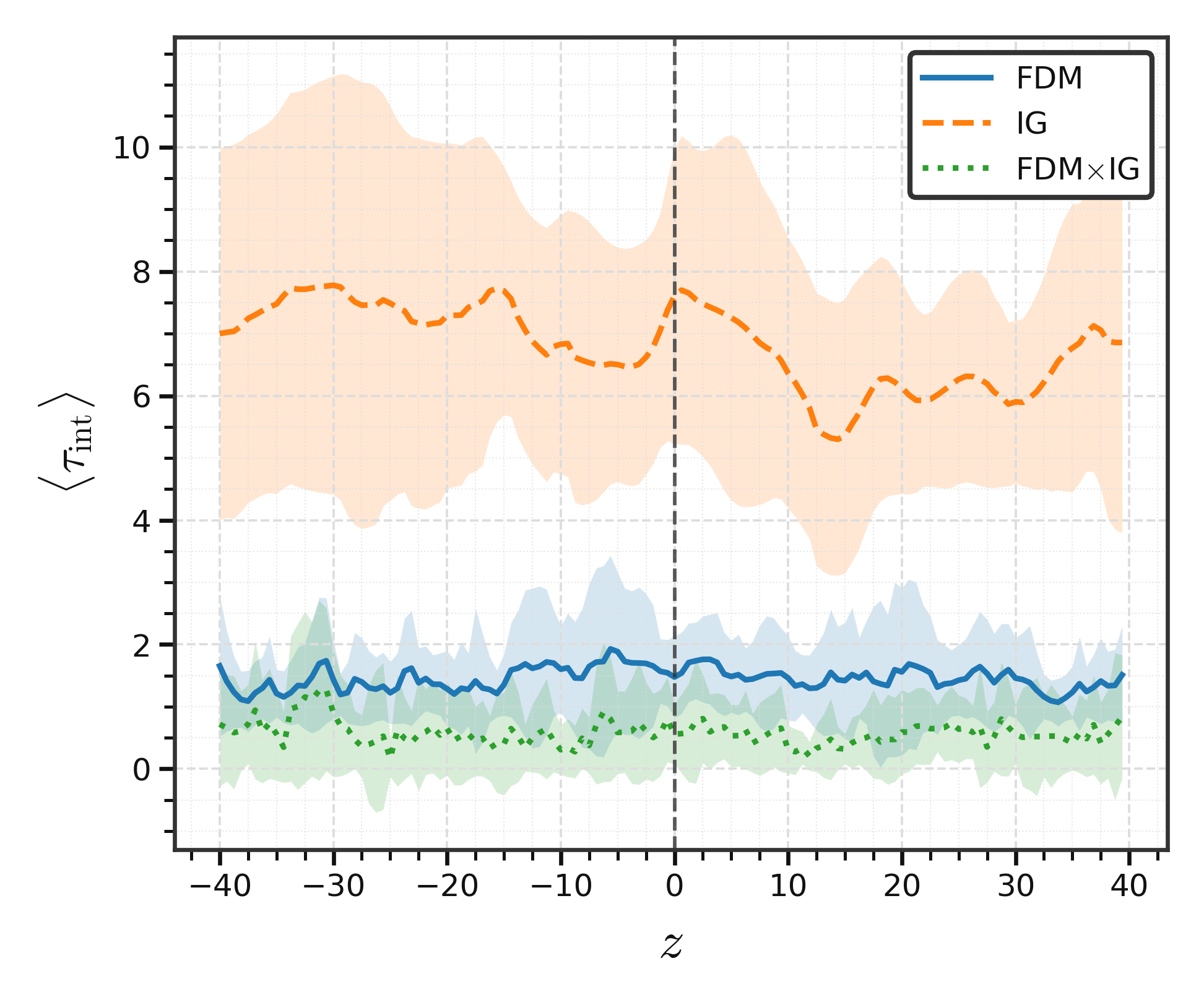}
\caption{Average over detector lines of the integrated correlation time $\langle \tau_{\mathrm{FDM}}(z) \rangle$ (solid), $\langle \tau_{\mathrm{IG}}(z) \rangle$ (dashed), and $\langle \tau_{\times}(z) \rangle$ (dotted) in the case $K=1$. Shaded regions indicate one standard deviation over the 16 detector lines.}
\label{fig:tau_int_mean_K1}
\end{figure}

{\it Statistical characterization.} To obtain a more robust characterization, we compute the mean integrated correlation time and its dispersion over the set of detectors. Figure~\ref{fig:tau_int_mean_K1} shows the averaged profiles $\langle \tau_{\mathrm{FDM}}(z) \rangle$, $\langle \tau_{\mathrm{IG}}(z) \rangle$, and $\langle \tau_{\times}(z) \rangle$, together with one standard deviation. 

The same qualitative behavior observed at the level of individual detectors is recovered in the detector-averaged quantities. The gas component exhibits the largest correlation times and the largest dispersion, reflecting its long-lived, coherent dynamics. In contrast, the FDM and cross-correlation signals remain both shorter and more clustered, indicating faster temporal decorrelation.

At the statistical level, an order of time scales can be identified:

\begin{equation}
\langle \tau_{\times} \rangle  ~ < ~
\langle \tau_{\mathrm{FDM}} \rangle ~<~
\langle \tau_{\mathrm{IG}} \rangle,
\end{equation}

\noindent which is consistently observed at all detector lines. This confirms that the rapid decorrelation of the FDM-gas coupling is not a local feature.

\begin{figure}
\centering
\includegraphics[width=8cm]{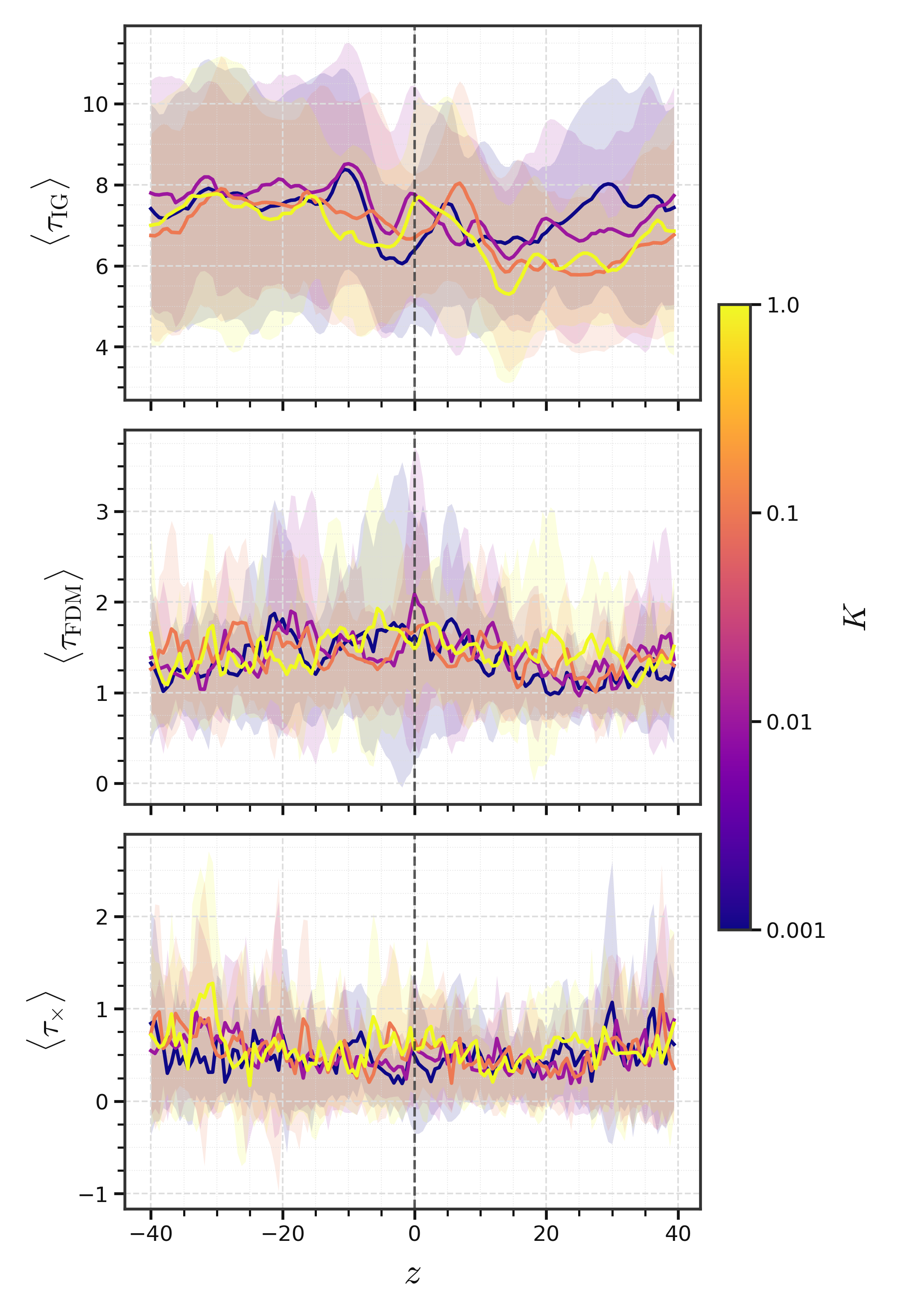}
\caption{Mean integrated correlation times as a function of $z$ for different values of the polytropic constant $K$. From top to bottom $\langle \tau_{\mathrm{IG}}(z) \rangle$, $\langle \tau_{\mathrm{FDM}}(z) \rangle$, and $\langle \tau_{\times}(z) \rangle$. The different colors correspond to values of $K$ spanning several orders of magnitude $K=10^{-3},10^{-2},10^{-1},1$. The profiles show only weak dependence on $K$, indicating that the temporal coherence properties of the system are dominated by the gravitationally driven dynamics rather than by the initial gas temperature.}
\label{fig:tau_int_K_scan}
\end{figure}

{\it Dependence on the initial polytropic constant.} To monitor the impact of the initial thermodynamic state of the gas, we compare the mean integrated correlation times for different values of the polytropic constant $K$. 
We thus repeat the analysis above using simulations with polytropic constants $K=10^{-3},10^{-2},10^{-1}$, in order to cover initially colder gas conditions and determine whether correlations change significantly.
Figure~\ref{fig:tau_int_K_scan} shows $\langle \tau_{\mathrm{IG}}(z) \rangle$, $\langle \tau_{\mathrm{FDM}}(z) \rangle$, and $\langle \tau_{\times}(z) \rangle$ for these values of $K$. Despite the significant variation in the initial gas temperature implied by these values, the resulting profiles remain remarkably similar.

In particular, the magnitude and spatial dependence of the correlation times show only minor variations across the explored parameter range. The gas component consistently exhibits the largest correlation times, the FDM remains in an intermediate regime, and the cross-correlation retains the shortest time scales.

This weak dependence on $K$ indicates that, once the system reaches a virialized core-halo configuration, in our simulations from $t=100$ to $t=200$, the temporal properties of the fluctuations are largely insensitive to the initial thermodynamic conditions of the gas. Instead, they appear to be primarily determined by the gravitational dynamics of the coupled system.

\section{Conclusions and discussion}
\label{sec:conclusions}

In this work we studied the temporal correlation properties of coupled Fuzzy Dark Matter (FDM) and baryonic matter configurations evolved with the Schr\"odinger-Poisson-Euler system. Starting from multimerger FDM initial conditions and random gas fluctuations, the system relaxed toward virialized core-halo structures whose dynamics were analyzed through temporal autocorrelations, cross-correlations, power spectra, and integrated correlation times.

Our analysis reveals a clear separation of temporal scales between the two components. The FDM density field exhibits rapidly varying fluctuations associated with interference and granularity, characterized by short coherence times and broad high-frequency spectra. In contrast, the gas evolves through smoother and more coherent modes dominated by low frequencies and longer correlation times.

The most important result of this work is that the temporal cross-correlation between the FDM and baryonic components remains weak and short-lived throughout the virialized configurations. Even though the gas dynamically reacts to the gravitational potential sourced by an FDM dominated halo, we do not observe strong temporal correlations associated with the fast interference-driven fluctuations of the granular FDM density field. This behavior persists across the explored range of initial gas temperatures. The cross-correlation is seen only at low frequencies, associated with the overall dynamics of the FDM, which in turn is dominated by the fundamental oscillation mode of the core \cite{Guzman2019}. We have seen that this correlation happens in a nearly perfect way when the configurations of FDM and IG oscillate around fermion-boson star solutions as demonstrated in \cite{AlvarezGuzman2023}, and also in the test field regime \cite{Widmark_2024}, however according to the results found here, it does not happen in the granular region.

The hierarchy of temporal scales, $\tau_{\times} < \tau_{\rm FDM} < \tau_{\rm IG}$, summarizes the dynamical behavior of the system, that is, the mutual FDM-gas correlations decorrelate faster than the intrinsic FDM fluctuations, while the gas evolves on significantly longer coherence times.

An interesting aspect of these results is that the characteristic time scale of the FDM fluctuations is not determined by the parabolic nature of the Schr\"odinger equation, but by wave interference and the oscillation modes of the virialized halo. In contrast, the gas has  hydrodynamic time scale associated with the sound speed. This different time scales explain the weak temporal coupling observed, the rapidly varying FDM fluctuations decorrelate before inducing a coherent response in the baryonic component.

These results suggest that baryonic matter primarily responds to the coarse-grained gravitational potential generated by the virialized halo rather than to the detailed phase-dependent structure of the FDM interference pattern. In particular, the rapid granular fluctuations characteristic of FDM do not appear to produce a strong temporal imprint on the gas distribution at the level of local density correlations.

Concerning the gas temperature. During the time-window of the analysis, when the structure is already virialized, for the various values of the polytropic constant, using a boson mass $m_B=10^{-22}$eV and domain side of 20 kpc,  the temperature of the ideal gas is of order $T\sim 1$ to $T=$1000 Kelvin, for $K=10^{-3}$ to $K=1$, respectively. These temperatures are well below the typical temperatures of hot interstellar gas ($T \sim 10^{6}$ Kelvin), although they lie within the cold neutral gas regime ($T\sim 10 - 10^4$ Kelvin), where pressure effects are expected to contribute less to the gas response to local fluctuation dynamics of the FDM, and we find that even in this regime the response to FDM granularity is small.

From a phenomenological perspective, our results indicate that direct observational signatures of FDM granularity through local baryonic temporal fluctuations may be weaker than expected from the dynamics of the common gravitational potential, even for a cold gas. At the same time, the existence of a well-defined hierarchy of temporal scales provides a quantitative characterization of the coupled dynamics between both components. In this sense, our work provides quantitative constraints on the extent to which interference-driven FDM dynamics may be observable through local baryonic temporal behavior, at least within the simple baryonic matter model assumed here.

Concerning the parameters of the FDM in our analysis, the scaling symmetry of the Schrödinger-Poisson system implies that the dimensionless solution analyzed here represents a family of physically scaled halos rather than a single physical case. Nevertheless, the present work considers only one dimensionless halo configuration, and it remains to be explored whether different dimensionless halo realizations or merger histories modify the temporal correlations reported here.

Despite the results, possible extensions of this analysis include the incorporation of additional baryonic physics, alternative equations of state, self-interactions, and more realistic galactic environments. It would also be relevant to study whether other observables, such as stellar kinematics or long-term secular effects, retain stronger signatures of FDM granularity.

\section*{Acknowledgments}
This research is supported by
SECIHTI Grant No. CFB-2025-I-759,
Laboratorio Nacional de C\'omputo de Alto Desempe\~no Grant No. 2026-8, and
CIC-UMSNH Grant No. 4.9.

\section*{Data availability}
The numerical results produced for this paper, 
as well as the data needed to reproduce all the plots are publicly available at \cite{datacorrelations}.

\bibliography{BECDM}

\end{document}